\documentclass[prl,floatfix,nofootinbib,superscriptaddress,showpacs,twocolumn,showkeys]{revtex4}
\usepackage{exscale}
\usepackage[intlimits]{amsmath}
\usepackage{amsfonts}
\usepackage{amssymb,amscd}
\usepackage[dvips]{epsfig}                   
\usepackage{array}
\usepackage{wrapfig}
\usepackage{bbm}
\usepackage{multirow}

\newcommand{\beqa}{\begin{eqnarray}}
\newcommand{\eeqa}{\end{eqnarray}}
\newcommand{\beq}{\begin{equation}}
\newcommand{\eeq}{\end{equation}}
\newcommand{\gS}[1]{#1\!\!\!\!\!\not~}

\newcommand{\pslash}{\gS{p}~}

\newcommand{\tr}{\textrm{tr}}

\newcommand{\ONE}{\mathbbm{1}}

\newcommand{\oursection}[1]{ {\vspace{0.2cm}\noindent \bf #1}\\}

\def\eq#1{(\ref{#1})}
\def\Eq#1{Eq.~(\ref{#1})}
 
\begin{document}

\title{Probing the gluon self-interaction in light mesons}

\author{Christian~S.~Fischer}
\affiliation{Institute for Nuclear Physics, 
 Darmstadt University of Technology, 
 Schlossgartenstra{\ss}e 9, 64289 Darmstadt, Germany}
\affiliation{GSI Helmholtzzentrum f\"ur Schwerionenforschung GmbH, 
  Planckstr. 1  D-64291 Darmstadt, Germany.}
\author{Richard Williams}
\affiliation{Institute for Nuclear Physics, 
 Darmstadt University of Technology, 
 Schlossgartenstra{\ss}e 9, 64289 Darmstadt, Germany}
\date{\today}

\begin{abstract}
We investigate masses and decay constants of light mesons from
a coupled system of Dyson--Schwinger and Bethe--Salpeter equations. 
We explicitly take into account dominant non-Abelian contributions
to the dressed quark-gluon vertex stemming from the gluon 
self-interaction. We construct the corresponding Bethe-Salpeter 
kernel that satisfies the axial-vector Ward-Takahashi identity.
Our numerical treatment fully includes all momentum dependencies
with all equations solved completely in the complex plane.
This approach goes well beyond the rainbow-ladder approximation 
and permits us to investigate the influence of the gluon 
self-interaction on the properties of mesons. As a first result 
we find indications of a nonperturbative cancellation of 
the gluon self-interaction contributions and pion cloud effects in 
the mass of the $\rho-$meson. 
\end{abstract}


\maketitle

{\noindent\bf Introduction}\\ 
Understanding the details of the light meson spectrum from 
underlying QCD is still an intricate and open problem of
considerable interest. Of course, pseudoscalar mesons are 
the (pseudo-)Goldstone bosons of QCD and as such enjoy an 
exceptional position amongst the hadronic states of QCD. 
Within the framework of Dyson--Schwinger (DSE) and 
Bethe-Salpeter (BSE) equations their Goldstone nature is 
retrieved in dynamical calculations provided constraints
from chiral symmetry, \emph{i.e.} the axial-vector Ward-Takahashi 
identity (axWTI), are taken into account, see \emph{e.g.}
\cite{Munczek:1994zz,Bender:1996bb}. However, apart from the 
masses of the (pseudo-)Goldstone bosons all other properties 
of light mesons such as masses, decay constants or charge radii are 
not governed by symmetry but depend on the details of the 
strong interaction of their constituents. This includes 
effects such as gluon self-interactions as well as pion cloud 
corrections \cite{Fischer:2008wy}. 

Within the DSE/BSE framework these effects are all contained 
in the structure of the nonperturbative quark-gluon vertex.
It is then clear that a simple rainbow-ladder parametrisation of 
the quark-gluon interaction in terms of vector couplings cannot 
be sufficient to describe the wealth of phenomena associated with
the internal structure of light mesons, see \emph{e.g.}~\cite{Eichmann:2008ae}
and references therein. In particular, a long-standing problem 
of rainbow-ladder is that it is too attractive, yielding masses of
$800$-$900$~MeV~\cite{Alkofer:2002bp} for the
axial-vector mesons.

Consequently there have been considerable efforts to go beyond
rainbow-ladder. Here, Abelian corrections to the quark-gluon vertex
have been considered in a number of works, see \emph{e.g.} 
\cite{Bender:1996bb,Bender:2002as,Bhagwat:2004hn,Watson:2004kd,Matevosyan:2006bk}.
These are, however, by far not the dominant contributions to the
vertex~\cite{Alkofer:2008tt}. Instead, genuinely 
non-Abelian diagrams including the gluon self-interaction are most
important. Up to now, these have only been considered on the level of 
the DSE for the quark-gluon vertex and the quark propagator 
~\cite{Alkofer:2008tt,Bhagwat:2004kj}.

Here we provide for a significant extension of these efforts. 
We present the first calculation of meson observables within a 
framework where the DSEs for the quark-gluon vertex and the 
quark propagator, as well as the BSE for mesons, have been coupled 
together without a trivialization of any momentum dependence.
Furthermore it is the first time that corrections from the dominant
non-Abelian part of the quark-gluon vertex including the gluon 
self-interaction are tested. 
We determine masses and decay constants of light mesons and 
evaluate the influence of the gluon self-interaction 
corrections as compared to the rainbow-ladder approximation. 

Our results are significant for $q\bar{q}$ bound states and 
therefore relevant for the vector mesons. In the axial-vector and 
scalar channels, we describe putative $q\bar{q}$ bound states which 
may or may not be realised in nature, see \emph{e.g.} \cite{Amsler:2004ps}. 
Nevertheless it is satisfactory that the inclusion of gluon 
self-interaction effects also in these channels leads to an 
improved description of light mesons.

\begin{figure}[t]
\begin{eqnarray}
\includegraphics[width=0.9\columnwidth]{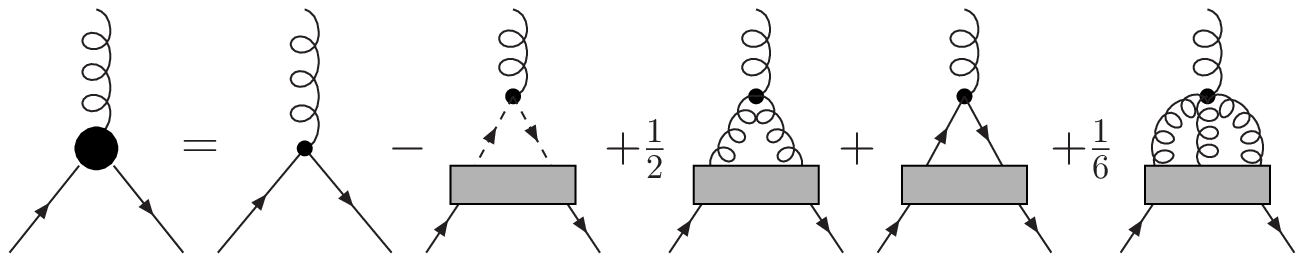}\label{dse1}\\
\includegraphics[width=0.9\columnwidth]{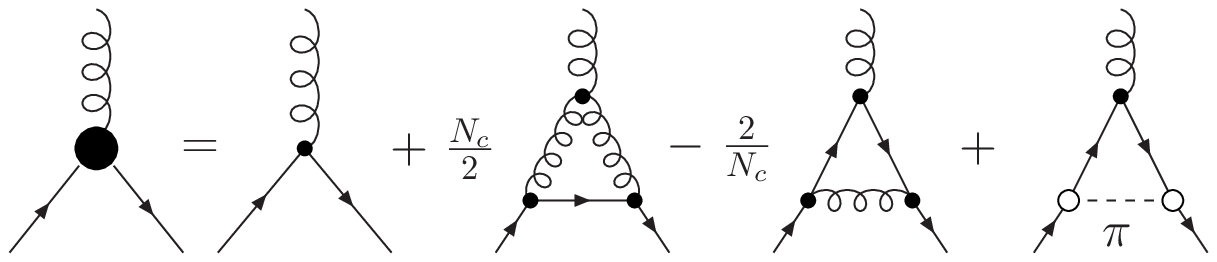}\label{dse2}
\end{eqnarray}
\caption{
The truncation employed for the quark-gluon vertex. All internal
propagators are dressed, with wiggly lines indicating gluons, straight
lines quarks and dashed lines mesons. White-filled circles indicate
bound-state amplitudes whilst black-filled represent vertex dressings.
Note that the last diagram is also proportional to $1/N_c$.
\label{fig:qgdse}}
\end{figure}

\oursection{The Quark-Gluon Vertex DSE}
The fundamental dynamical quantity in the 
DSE/BSE approach to hadrons is the quark-gluon
vertex, whose specification will ultimately determine the truncation
scheme. We therefore begin with a discussion of its DSE depicted 
in \Eq{dse1} and \Eq{dse2} of Fig.~\ref{fig:qgdse}. 
Following the detailed investigation of the quark-gluon vertex
in Refs.~\cite{Alkofer:2008tt,Fischer:2007ze} we approximate the full
DSE \eq{dse1} with the (nonperturbative) one-loop structure of \eq{dse2}. 
Here the first `non-Abelian' loop-diagram in \eq{dse2} subsumes the first 
two diagrams in the full
DSE to first order in a skeleton expansion of the four-point functions.
The two-loop diagram in the full DSE \eq{dse1} is neglected. This 
approximation is well justified in the large and small momentum r\`egime
\cite{Alkofer:2008tt,Bhagwat:2004kj} and is assumed to be tractable for
intermediate momenta. The remaining `Abelian' contributions are split
into the non-resonant second loop-diagram in \eq{dse2} and a third
diagram containing effects due to hadron backreactions.
These are dominated by the lightest hadrons, \emph{i.e.} pseudoscalar 
mesons.

Note that the non-Abelian and non-resonant Abelian diagrams are associated 
with colour factors $N_c/2$ and $-2/N_c$, respectively. The diagram 
including the pion exchange is also 
proportional to $1/N_c$ due to the implicit $1/\sqrt{N_c}$-dependence of the
two pion amplitudes. Thus, both Abelian diagrams are suppressed by a 
factor of $N_c^2$ as compared to the non-Abelian one, a fact that is 
also evidenced through direct numerical calculation
\cite{Alkofer:2008tt}. 
Consequently, these diagrams are by far not the leading ones in the 
vertex-DSE.  
Moreover, as concerns meson masses, the Abelian diagrams
are generally attractive. The effects of these diagrams on the mass 
of the $\rho$-meson have been estimated to be about $30$~MeV due to 
non-resonant Abelian diagrams in the quark-gluon vertex \cite{Watson:2004kd} 
with another $90$~MeV due to pion cloud effects \cite{Fischer:2008wy}.

\begin{figure}[b]
\centerline{\includegraphics[width=0.8\columnwidth]{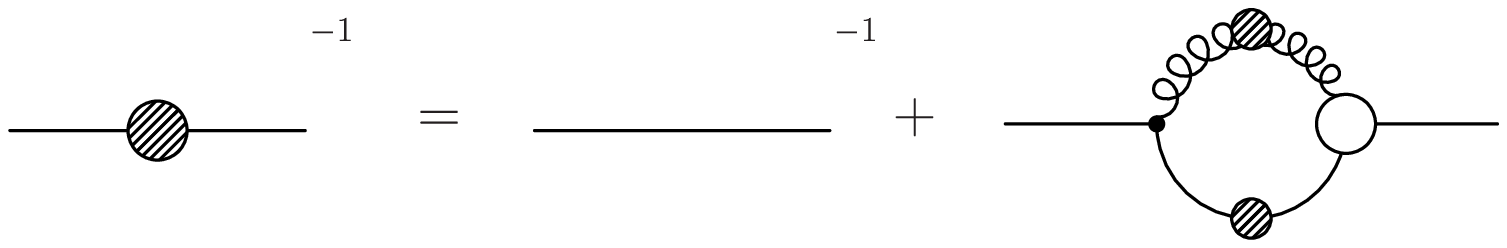}}
\caption{The DSE for the fully dressed
quark propagator.}\label{fig:quarkdse}
\end{figure}

In this letter we concentrate on the leading non-Abelian
diagram in \Eq{dse2} and explore its effects on meson observables
as compared to pure rainbow-ladder
approximations. In order to keep our calculation tractable we employ
the well-established strategy of absorbing all internal vertex dressings
into effective dressing functions for the two
internal gluon propagators. The resulting DSE for the 
quark-gluon vertex $\Gamma^{\mu}(p_1,p_2)$ with quark momenta $p_1$ 
and $p_2$ and gluon momentum $p_3$ reads 
%
\begin{eqnarray} 
  \Gamma^{\mu}(p_1,p_2) &=& Z_{1F} \gamma^\mu+ 
\left( \frac{-i N_c}{2} g^2 Z_{1F}^2 Z_1\right) \times \\ 
&&\hspace{-40pt}\times \int_q 
\left\{  \gamma^\nu S(q)
\gamma^\rho 
\Gamma^{\rm 3g}_{\sigma\theta\mu}(k_1,k_2)
D_{\nu\sigma}(k_1) D_{\rho\theta}(k_2)
\right\}
\nonumber\label{eqn:qg}
\end{eqnarray}
%
with $\int_q \equiv \int \frac{d^4q}{(2\pi)^4}$, the renormalization 
factors $Z_{1F}$,$Z_1$ and 
$\Gamma^{\rm 3g}$ the bare three-gluon vertex.  
In general, the quark-gluon vertex $\Gamma^\mu$ is given by a 
combination of twelve independent tensors built up of the 
quark momenta $p_1^\mu$, $p_2^\mu$ and $\gamma^\mu$; for
a detailed discussion see Ref.~\cite{Alkofer:2008tt}. 
The dressed quark and gluon propagators are given by 
\begin{eqnarray}
S(p)&=& [-i\, \pslash\! A(p^2) + B(p^2)]^{-1} \,,
\label{quark}\\
D_{\mu\nu}(k) &=&  \left(\delta_{\mu \nu} 
- \frac{k_\mu k_\nu}{k^2}\right)                                        
\frac{Z(k^2)}{k^2}\,, \label{gluon}
\end{eqnarray}
where $Z$ is the gluon dressing.
The quark dressing functions $A$, $B$ are determined from the DSE for the quark 
propagator given diagrammatically in Fig.~\ref{fig:quarkdse}.
With bare quark mass $m$ and renormalization factor $Z_2$ it reads 
\begin{eqnarray}
  S^{-1}(p) &=& Z_{2} [-i\, \pslash\! + m]\nonumber\\[-2mm] \label{DSE1}\\[-2mm]
  +&&\hspace{-15pt} g^{2}C_{F}Z_{1F}\int_q\,  \gamma_{\mu}S(q)\Gamma_{\nu}(q,k)
  D_{\mu\nu}(k)   \,.\nonumber \label{eqn:quarkdse}
\end{eqnarray}

What remains to be specified in both the vertex and the quark DSE
is the effective dressing of the gluon propagator. Here we use 
the momentum dependent ansatz~\cite{Alkofer:2002bp}
\begin{equation}
  Z(q^2) = \frac{4\pi}{g^2}\frac{\pi D }{\omega^2}\;q^4\;  e^{-q^2/\omega^2}\;,\label{eqn:gluon}
\end{equation}
with two parameters $D$ and $\omega$ which provide for the scale and 
strength of the effective gluon interaction. Naturally, such an ansatz
provides only a first step towards a full calculation of the non-Abelian
diagram including input from the DSEs for the three-gluon vertex and 
the gluon propagator. Given that the numerical treatment of the coupled system
of vertex-DSE, quark-DSE and the Bethe-Salpeter equation for light mesons
is somewhat involved we defer such a calculation for future work. Nevertheless
we believe that the ansatz \eq{eqn:gluon} is sufficient to provide for reliable 
qualitative results as concerns the effects due to the non-Abelian diagram onto 
meson properties. In particular it is not sensitive to the question
of scaling vs. decoupling \cite{Fischer:2008uz} in the deep infrared, $p<50$ MeV: 
both scaling and decoupling 
lead to a combination of three-gluon vertex and gluon propagator dressings 
that is vanishing in the infrared in qualitative agreement with the 
ansatz \eq{eqn:gluon}. In addition, quantitative effects in the interaction
below $p<50$ MeV are not expected to affect observables in the flavour
non-singlet sector since 
the dynamical mass of the quark, $M \approx 350$ MeV, suppresses all
physics on scales $p \ll M$ (see however \cite{Alkofer:2008et}). Finally,
the ansatz \eq{eqn:gluon} is not sensitive to details of the Slavnov-Taylor 
identity (STI) for the three-gluon vertex: exactly those longitudinal parts 
of the vertex that are constrained from the STI are projected out in any Landau
gauge calculation by the attached transverse gluon propagators.

As concerns our numerical treatment we solve the coupled system of quark-gluon vertex 
DSE and quark DSE for complex Euclidean momenta $p_1$ and $p_2$ with 
the `shell-method' described in the second work of 
Ref.~\cite{Fischer:2007ze}. Through judicious choice of momentum routing in 
both the quark DSE and vertex DSE, this can be accomplished without 
unconstrained analytic continuation of the gluon propagator and three-gluon 
vertex. The BSE is solved for complex momenta by standard methods
\cite{Alkofer:2002bp,Maris:1997tm}.

\begin{figure}[h]
\begin{eqnarray}
	\begin{array}{c}
	\includegraphics[scale=0.4]{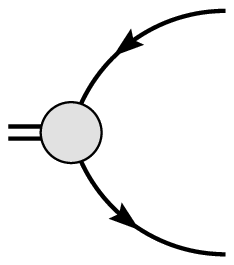}\\
	\end{array}
	&=& 
	\begin{array}{c}
	\includegraphics[scale=0.4]{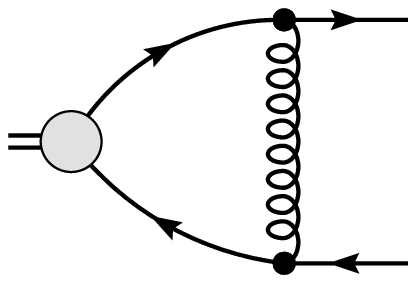}\\
	\end{array}
	+
	\begin{array}{c}
	\includegraphics[scale=0.4]{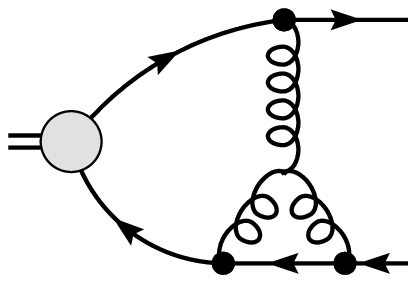}\\
	\end{array}
	+
	\begin{array}{c}
	\includegraphics[scale=0.4]{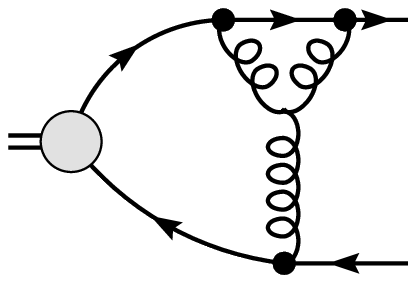}\\
	\end{array}
	\nonumber
	\end{eqnarray}
\caption{The axWTI preserving BSE corresponding to our vertex truncation. All 
propagators are dressed, with wiggly and straight lines showing gluons
and quarks respectively.\label{fig:ourbse}}
\end{figure}

\oursection{The Bethe-Salpeter equation}
The Bethe-Salpeter amplitude $\Gamma(p;P) \equiv \Gamma^{(\mu)}(p;P)$ for
a bound state of mass $M$ is calculated through
\begin{equation}
  \left[\Gamma(p;P)\right]_{tu} = \lambda \, \int_k
K_{tu}^{rs}(p,k;P)\left[S(k_+)\Gamma(k;P)S(k_-)\right]_{sr}\,,
\label{eqn:bse2}
\end{equation}
with $k_\pm = k \pm P/2$ and eigenvalue $\lambda(P^2=-M^2)=1$.
The amplitude can be decomposed into at most eight Lorentz and Dirac structures 
constrained by transformation properties under CPT~\cite{LlewellynSmith:1969az}. 
It is well-known that one may construct a Bethe-Salpeter kernel $K_{tu}^{rs}$
satisfying the axWTI by means of a functional derivative of the quark 
self-energy \cite{Bender:1996bb,Munczek:1994zz,Maris:1997tm,Maris:2005tt}. 
Applying this cutting procedure to the quark DSE specified by
Eq.~(\ref{DSE1}) with the quark-gluon vertex of Eq.~(\ref{eqn:qg})
yields the Bethe-Salpeter equation portrayed in Fig.~\ref{fig:ourbse} \cite{Maris:2005tt}.

\begin{figure}[b!]
  \begin{center}
\begin{eqnarray}
	\ONE	=  2\frac{\partial}{\partial P^2} \tr\!\!\!&\Bigg[&\!\!\!
	\begin{array}{c}
	\includegraphics[scale=0.4]{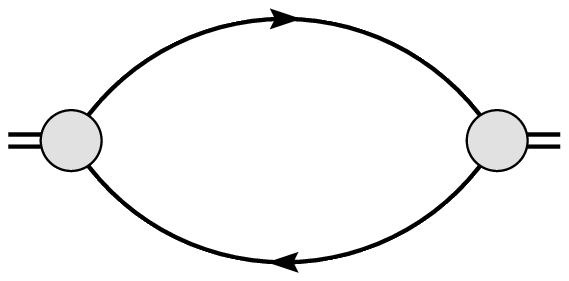}\\
	\end{array}
	+
	\begin{array}{c}
	\includegraphics[scale=0.4]{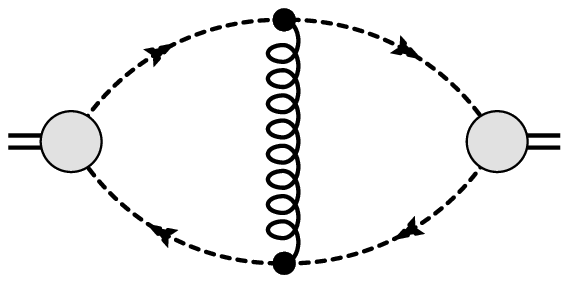}\\
	\end{array}
	\nonumber\\
	\!\!\!&+&\!\!\!
	\begin{array}{c}
	\includegraphics[scale=0.4]{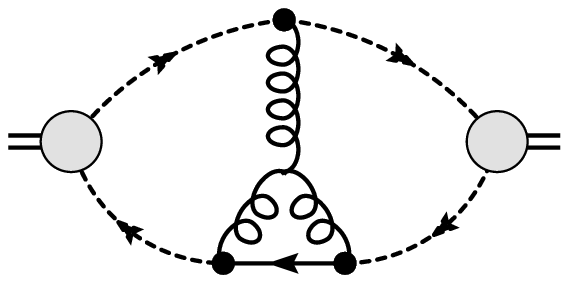}\\
	\end{array}
	+
	\begin{array}{c}
	\includegraphics[scale=0.4]{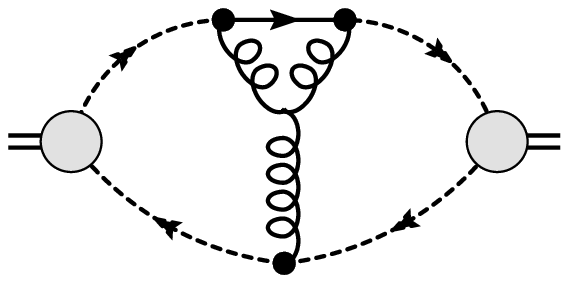}\\
	\end{array}
\Bigg]\nonumber
	\end{eqnarray}
\caption{Normalisation of the Bethe-Salpeter amplitude. Dashed lines
represent quark propagators that are kept fixed under action of the
derivative.\label{fig:norm}}
\end{center}
\end{figure}

\oursection{\bf Normalisation}
Since we solve the homogeneous Bethe-Salpeter equation, the correct 
normalisation of the amplitude is achieved through an auxiliary condition~\cite{Cutkosky:1964zz}, 
given pictorially in Fig.~\ref{fig:norm}. This involves evaluating three-loop
integrals  over nonperturbative quantities, which we tackle with the aid of
standard Monte-Carlo techniques. 
Alternatively, using the eigenvalue $\lambda$ in \eq{eqn:bse2} and
the conjugate amplitude $\overline{\Gamma}(k,-P) = C\Gamma(-k,-P)C^{-1}$ one may use the 
equivalent normalisation condition~\cite{Nakanishi:1965zza} 
\begin{equation}\label{eqn:norm}
\left( \frac{d\ln(\lambda)}{d P^2}
  \right)^{-1} \!\!\!=\tr\int_k \,3\, 
\overline{\Gamma}(k,-P)  S(k_+)\Gamma(k,P)S(k_-)\,.
\end{equation}
This requires significantly less numerical effort and can be simply applied to
all truncations of the BSE.

We will see that the diagrams of Fig.~\ref{fig:norm} beyond the impulse 
approximation give large corrections of the order of $30\%$. This is important for observables calculated from the
amplitudes such as leptonic decay constants~\cite{Maris:1997tm} 

\begin{table*}[t!]
\renewcommand{\tabcolsep}{0.5pc} 
\renewcommand{\arraystretch}{1.1} 
\begin{tabular}{@{}c||cc|ccc|ccc|ccc}
  Model & $\omega$ & $D$ & $m_\pi$ & $f_\pi$ & $\hat{f_\pi}$ &
$m_\rho$  & $f_\rho$ & $\hat{f_\rho}$ & $m_\sigma$ & $m_{a_1}$ &
$m_{b_1}$\\
  \hline\hline
R-L& \multirow{2}{*}{$0.50$} & \multirow{2}{*}{$16$} & $138$  & $94$ 
& --      &  $758$  & $154$ & --       & $645$ & $926$ & $912$\\
  BTR   &        &      & $138$  & $111$  & $(127)$ &  $881$  & $176$
&  $(181)$ & $884$ & $1055$ & $972$\\
  \hline
  R-L   & \multirow{2}{*}{$0.48$} & \multirow{2}{*}{$20$} & $138$ &
$95$   & --      &  $763$  & $154$ & --       & $676$ & $937$ & $895$\\
  BTR   &        &      & $141$ & $112$  & $(129)$ &  $887$  & $176$
&  $(181)$ & $886$ & $1040$ & $946$\\
  \hline
  R-L   & \multirow{2}{*}{$0.45$} & \multirow{2}{*}{$25$} & $136$ &
$92$   & --      &  $746$  & $149$ & --       & $675$ & $917$ & $858$\\
  BTR   &        &      & $142$ & $110$  & $(128)$ &  $873$  & $173$
&  $(177)$ & $796$ & $1006$ & $902$\\
  \hline\hline
Experiment~\cite{Amsler:2008zzb} &     &      & $138$  & $92.4$ &  --     & $776$ & $156$ &
-- & $400$--$1200$ & $1230$ & $1230$ \\
\hline
\end{tabular}
\caption{Masses and decay constants for a variety of mesons, calculated
using rainbow-ladder (R-L) and our beyond-the-rainbow (BTR) truncation.
Decay constants are determined from the full normalisation condition,
except those indicated by a caret where only the leading term of the
impulse approximation is used. Masses and decay constants are given in
MeV. \label{results}}
\end{table*}

\oursection{Results}
Table~\ref{results} details the results of our truncation scheme,
including gluon self-interaction effects in all twelve tensor 
structures of the quark-gluon vertex, compared to the rainbow-ladder 
with only vector-vector interactions. The model parameters 
$\omega$ and $D$ were tuned such that for the latter we obtain 
reasonable pion observables, and the quark mass is fixed at $5$~MeV.
We do not fit observables for our beyond rainbow-ladder truncation scheme since the inclusion
of additional resonant and non-resonant corrections would require
further parameter tuning.

In all cases the non-Abelian corrections are 
seen to have only a small impact on the mass of the pion, though 
the decay constant is enhanced from $94$~MeV to about $110$~MeV. 
This shift is comparable in size to the negative one of the order of 
$10$~MeV \cite{Fischer:2008wy} induced by pion cloud effects,
\emph{i.e.} the third loop-diagram in \eq{dse2}. We also
investigated the impact of the Abelian, second loop-diagram in 
\eq{dse2} without use of the real-axis approximation made in 
\cite{Watson:2004kd}. We confirm that the change in the pion mass 
is negligible for their preferred value of $G=0.5$ and find a small reduction of $\sim2$~MeV in the decay constant.
Once all corrections are combined, we expect significant 
cancellations such that the mass and decay constant of the pion
are close to the experimental value.

\begin{figure}[b!]
\includegraphics[width=0.8\columnwidth]{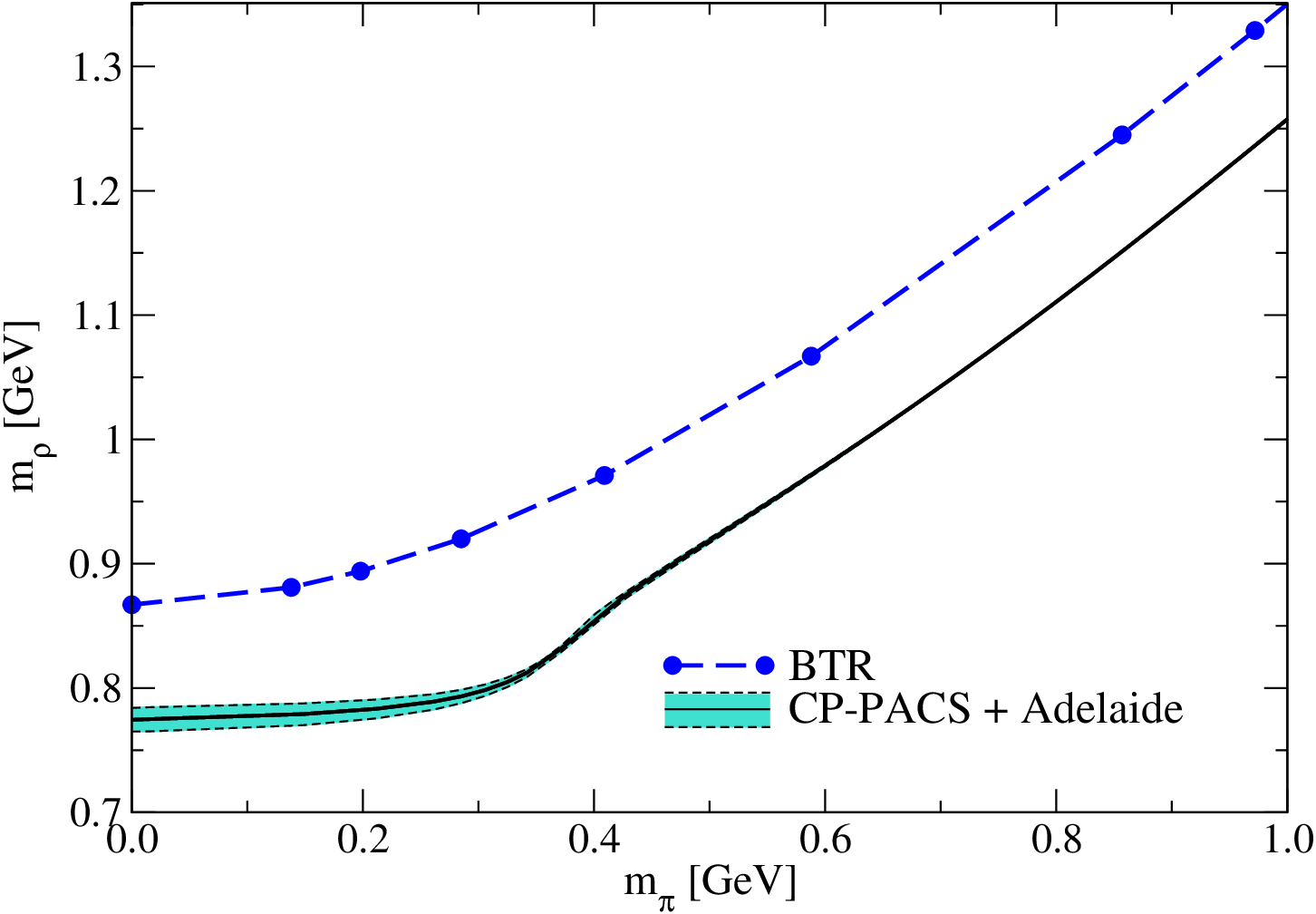}\label{rhopi}\\
\caption{
Rho mass as a function of the pion mass compared to a
chiral extrapolation based on (corrected) lattice data, 
Ref.~\cite{Allton:2005fb}.
\label{fig:rhopi}}
\end{figure}

For the rho meson, including the gluon self-interaction enhances its 
mass by $\sim120$~MeV compared to pure rainbow-ladder, with the 
decay constant increased by $\sim20$~MeV. This is an intriguing 
result since it has long been suspected that corrections beyond 
rainbow-ladder cancel amongst themselves in the vector channel 
\cite{Bender:1996bb,Bender:2002as}. Indeed, the resonant and non-resonant 
contributions from the Abelian diagrams in \eq{dse2} are known 
to provide reductions of the rho mass of $\sim90$~MeV and $30$~MeV 
respectively~\cite{Fischer:2008wy,Watson:2004kd}. Similar cancellations 
happen for the decay constant. We therefore 
see for the first time strong evidence of this nonperturbative 
cancellation mechanism. This is one of the main results of this letter.

Fig.~\ref{fig:rhopi} shows the rho mass as a function of the pion
mass for our truncation beyond-the-rainbow (BTR) compared to a
chiral extrapolation based on (corrected) lattice data, 
Ref.~\cite{Allton:2005fb}. Due to the discussion above we expect 
the explicit inclusion of the resonant and non-resonant Abelian 
contributions in our BTR-scheme to move our results close to the 
lattice data. 

For completeness, we also report the masses of the scalar and
axial-vectors in Table~\ref{results}. In all cases we see an 
enhancement compared to rainbow-ladder. 

\oursection{Conclusions}
We presented an exploratory study of light mesons using a sophisticated
truncation of the Bethe-Salpeter equations beyond rainbow-ladder, in
which we consider the gluon self-interaction contributions to the
quark-gluon vertex. Close to the chiral limit we obtain masses for 
the rho meson of $\sim900$~MeV, consistent with extrapolations from 
quenched lattice simulations. There is evidence that the subsequent inclusion
of pion cloud effects and non-resonant contributions to the 
quark-gluon interaction brings the rho mass back to its physical value 
thus supporting a long suspected nonperturbative cancellation mechanism. 
Our truncation provides for a well-founded setup to further explore the
details of the nonperturbative quark-gluon interaction and gluon 
self-interaction effects in mesons. In this respect it is complementary 
to the one very recently suggested in Ref.~\cite{Chang:2009zb}.

\oursection{Acknowledgements}
We thank Pete Watson for discussions.
This work was supported by the Helmholtz
Young Investigator Grant VH-NG-332 and the
Helmholtz International Center for FAIR
within the LOEWE program of the State of Hesse.


\begin{thebibliography}{99}

\bibitem{Bender:1996bb}
  A.~Bender, C.~D.~Roberts and L.~Von Smekal,
  Phys.\ Lett.\  B {\bf 380}, 7 (1996).

\bibitem{Munczek:1994zz}
  H.~J.~Munczek,
  Phys.\ Rev.\  D {\bf 52} (1995) 4736.

\bibitem{Fischer:2008wy}
  C.~S.~Fischer and R.~Williams,
  PRD {\bf 78} (2008) 074006.

\bibitem{Eichmann:2008ae}
  G.~Eichmann {\emph et al.}, 
  Phys.\ Rev.\  C {\bf 77} (2008) 042202.

\bibitem{Alkofer:2002bp}
  R.~Alkofer, P.~Watson and H.~Weigel,
  Phys.\ Rev.\  D {\bf 65} (2002) 094026.


\bibitem{Bender:2002as}
  A.~Bender {\emph et al.}, 
  Phys.\ Rev.\  C {\bf 65}, 065203 (2002);

\bibitem{Bhagwat:2004hn}
  M.~S.~Bhagwat {\it et al.}
  Phys.\ Rev.\  C {\bf 70} (2004) 035205.

\bibitem{Watson:2004kd}
  P.~Watson, W.~Cassing and P.~C.~Tandy,
  Few Body Syst.\  {\bf 35} (2004) 129;
  P.~Watson and W.~Cassing,
  Few Body Syst.\  {\bf 35} (2004) 99.

\bibitem{Matevosyan:2006bk}
  H.~H.~Matevosyan, A.~W.~Thomas and P.~C.~Tandy,
  Phys.\ Rev.\  C {\bf 75} (2007) 045201.


\bibitem{Alkofer:2008tt}
  R.~Alkofer {\emph et al.}, 
  Annals Phys.\  {\bf 324}, 106 (2009).

\bibitem{Bhagwat:2004kj}
  M.~S.~Bhagwat and P.~C.~Tandy,
  Phys.\ Rev.\  D {\bf 70} (2004) 094039.

\bibitem{Amsler:2004ps}
  C.~Amsler and N.~A.~Tornqvist,
  Phys.\ Rept.\  {\bf 389} (2004) 61;
  M.~Wagner and S.~Leupold,
  Phys.\ Rev.\  D {\bf 78} (2008) 053001.


\bibitem{Fischer:2007ze}
  C.~S.~Fischer, D.~Nickel and J.~Wambach,
  Phys.\ Rev.\  D {\bf 76} (2007) 094009;
  C.~S.~Fischer, D.~Nickel and R.~Williams,
  Eur.\ Phys.\ J.\  C {\bf 60}, 1434 (2008).

\bibitem{Fischer:2008uz}
  C.~S.~Fischer, A.~Maas and J.~M.~Pawlowski,
  Annals Phys. in press; arXiv:0810.1987, and Refs. therein.

\bibitem{Alkofer:2008et}
  R.~Alkofer, C.~S.~Fischer and R.~Williams,
  Eur.\ Phys.\ J.\  A {\bf 38}, 53 (2008).

\bibitem{LlewellynSmith:1969az}
  C.~H.~Llewellyn-Smith,
  Annals Phys.\  {\bf 53} (1969) 521.

\bibitem{Maris:1997tm}
  P.~Maris and C.~D.~Roberts,
  PRC {\bf 56} (1997) 3369.

\bibitem{Maris:2005tt}
  P.~Maris and P.~C.~Tandy,
  Nucl.\ Phys.\ Proc.\ Suppl.\  {\bf 161} (2006) 136.


\bibitem{Cutkosky:1964zz}
  R.~E.~Cutkosky and M.~Leon,
  Phys.\ Rev.\  {\bf 135} (1964) B1445.


\bibitem{Nakanishi:1965zza}
  N.~Nakanishi,
  Phys.\ Rev.\  {\bf 138} (1965) B1182.

\bibitem{Allton:2005fb}
  C.~R.~Allton {\em et al.}, 
  Phys.\ Lett.\  B {\bf 628} (2005) 125.

\bibitem{Amsler:2008zzb}
  C.~Amsler {\it et al.}  [PDG],
  Phys.\ Lett.\  B {\bf 667} (2008) 1.


\bibitem{Chang:2009zb}
  L.~Chang and C.~D.~Roberts,
  Phys.\ Rev.\ Lett.\  {\bf 103} (2009) 081601.

    
\end{thebibliography}
\end{document}